\documentclass[aps,prl,twocolumn,superscriptaddress]{revtex4-2}

\usepackage{orcidlink}
\usepackage{graphicx}%
\usepackage{multirow}%
\usepackage{amsmath,amssymb,amsfonts}%
\usepackage{bm}%
\usepackage{amsthm,ulem}%
\usepackage{mathrsfs}%
\usepackage[version=4]{mhchem}
\usepackage[title]{appendix}%
\usepackage{xcolor}%
\usepackage{textcomp}%
\usepackage{booktabs}%
\usepackage{algorithm}%
\usepackage{algorithmicx}%
\usepackage{algpseudocode}%
\usepackage{listings}%
\usepackage[separate-uncertainty=true, multi-part-units=single]{siunitx}
\usepackage{import}
\usepackage{hyperref}
\hypersetup{colorlinks,breaklinks,
	urlcolor=blue,
	linkcolor=blue,
	citecolor=blue}

\begin{document}


\title{Interplay between timescales governs the residual activity of a harmonically bound active Brownian particle}

\author{Sanatan Halder \orcidlink{0009-0002-2457-0449}}
\email[Contact author: ]{sanatanh@iitk.ac.in}
\affiliation{Department of Physics, Indian Institute of Technology Kanpur, Kanpur - 208016, India}

\author{Manas Khan \orcidlink{0000-0001-6446-3205}}
\email[Contact author: ]{mkhan@iitk.ac.in}
\affiliation{Department of Physics, Indian Institute of Technology Kanpur, Kanpur - 208016, India}


\begin{abstract}
Active microparticles in confining potentials manifest many complex dynamical phenomena as their activity competes with confinement. The steady-state position distributions of harmonically bound active Brownian particles exhibit a crossover from Boltzmann-like to bimodal, commonly referred to as passive and active regimes, respectively, with variations in activity and confinement strength. By studying optically trapped active Janus colloids, numerical simulations, and analytical calculations, we demonstrate that the underlying crossover is from activity-dominated bound dynamics to activity-depleted Brownian motion in a radially displaced harmonic well; and is solely governed by the ratio of the persistence time to the equilibration time in the harmonic potential, being independent of the propulsion speed. 
\end{abstract}

\maketitle

\paragraph{Introduction.}
The active dynamics of self-propelled particles have been a major frontier of contemporary research because of their relevance in describing fundamental nonequilibrium processes and potential applications \cite{Ramaswamy2010, Romanczuk2012, Marchetti2013, Cates2015, Bechinger2016, Fodor2018}. They are most commonly realized using phoretically propelled Janus microspheres \cite{howseSelfMotileColloidalParticles2007,jiangActiveMotionJanus2010, Volpe2011} and are modeled as \textit{active Brownian particles} (ABPs) \cite{Romanczuk2012, Bechinger2016, Fodor2018, basuActiveBrownianMotion2018}. Confining potentials or physical restrictions on their motion give rise to intriguing nonequilibrium phenomena, \textit{e.g.}, accumulation toward the periphery \cite{Berke2008, elgetiWallAccumulationSelfpropelled2013, leeActiveParticlesConfinement2013}, anomalous sedimentation \cite{solonActiveBrownianParticles2015, Ginot2015}, self-induced polar ordering \cite{bauerleFormationStableResponsive2020, lavergneGroupFormationCohesion2019, hennesSelfInducedPolarOrder2014}, and active glasses in dense systems \cite{Ni2013, Berthier2013, Janssen2019}, as activity competes with confinement owing to narrow channels, crowded, porous, and viscoelastic environments \cite{Ni2013, Berthier2013, filyDynamicsSelfpropelledParticles2014, BenIsaac2015, Bechinger2016, Ribeiro2018, narinderActiveParticlesGeometrically2019, Janssen2019, Sprenger2022, Moore2023}. \textit{Harmonically bound ABPs} (HBABPs) provide an excellent means of modeling and understanding these phenomena.

In recent years, there has been a surge in studies on HBABPs, mostly employing analytical calculations  \cite{pototskyActiveBrownianParticles2012, basuLongtimePositionDistribution2019, chaudhuriActiveBrownianParticle2021,malakarSteadyStateActive2020, Santra2021, buttinoniActiveColloidsHarmonic2022, caraglioAnalyticSolutionActive2022, Baldovin2023} and numerical simulations \cite{pototskyActiveBrownianParticles2012, malakarSteadyStateActive2020, Santra2021, buttinoniActiveColloidsHarmonic2022}, along with a few experiments \cite{takatoriAcousticTrappingActive2016, schmidtNonequilibriumPropertiesActive2021, buttinoniActiveColloidsHarmonic2022}, exploring a crossover where the steady-state position distribution changes from Boltzmann-like to bimodal with varying strengths of activity against confinement. Hence, it is commonly described as a passive to active or equilibrium to strongly-nonequilibrium transition. However, identifying the crossover based on the form of the steady-state position distribution, which contradicts other statistical descriptions of HBABP dynamics \cite{halder2025PRE}, remains debatable. Therefore, detailed dynamical analyses to reliably define and measure the signature of activity in HBABP dynamics are essential for comprehending this crossover. Although a few studies have investigated and utilized the orbital dynamics of the HBABP or analogous inertial systems \cite{zongOpticallyDrivenBistable2015, Schmidt2018, dauchotDynamicsSelfPropelledParticle2019, brontecirizaOpticallyDrivenJanus2023, Patil2021}, the manifestation of activity in the resultant motion has not been explored.

\begin{figure}[hb]
	\centering
	\includegraphics[width = 0.85\linewidth]{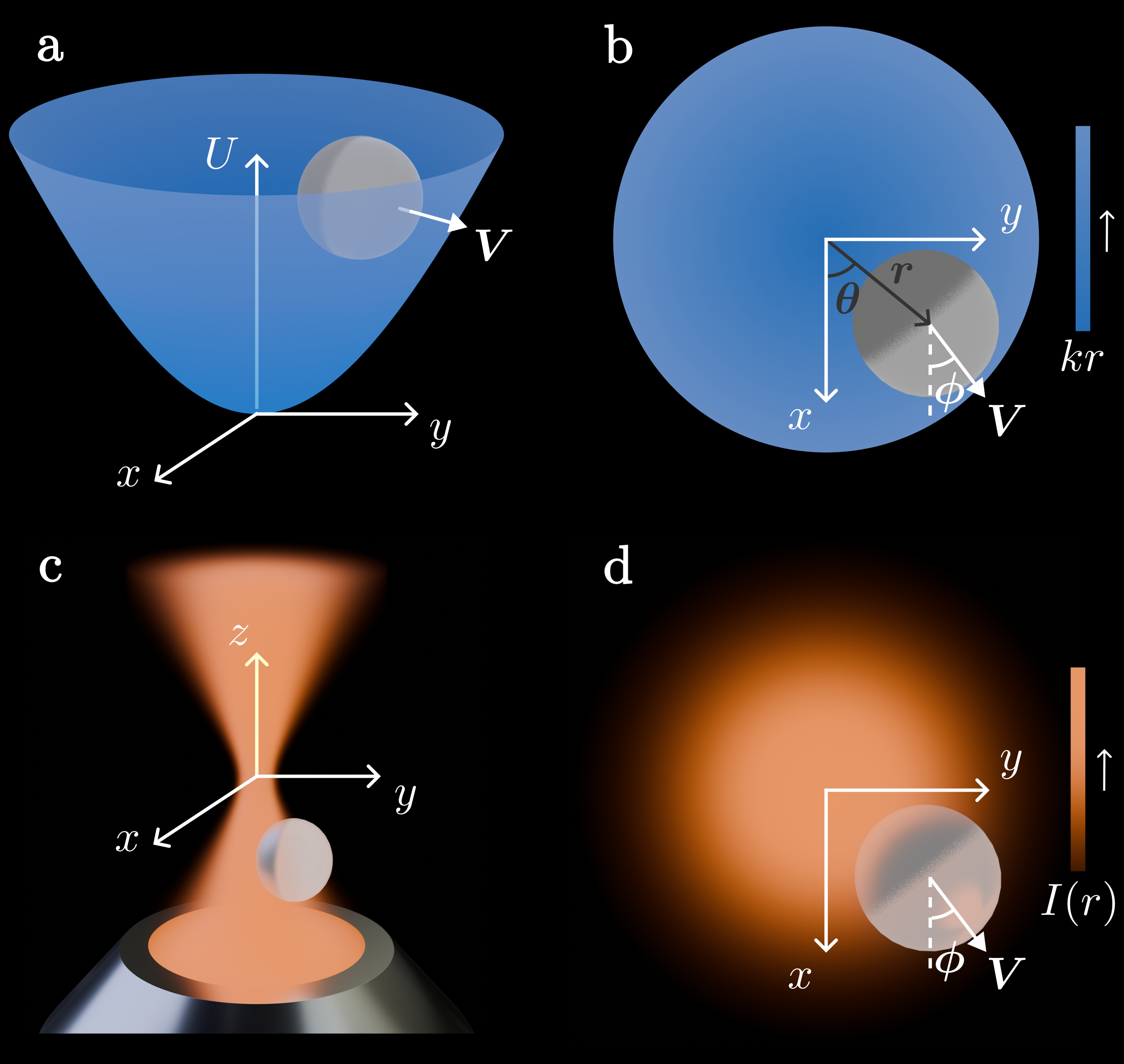}
	\caption{Schematic description of an HBABP (top row) and its experimental realization by a phoretically active Janus colloid in an optical trap (bottom row). (a) An ABP (dark grey - light grey sphere) with propulsion $\bm{V}$ is confined in a 2D harmonic potential (blue). (b) The radial variation in the corresponding restoring force field ($ - k \bm{r}$) is shown as a blue gradient, where $\bm{V}$ makes an angle $\phi (t)$ with the $x$-axis. $r (t)$ and $\theta (t)$ represent the plane-polar coordinates. (c) Optical trapping of a Janus colloid using a tightly focused laser beam (orange). (d) Gaussian intensity profile ($I (r)$, orange gradient) is shown along with the active Janus particle.
		\label{fig:HBABP}}
\end{figure}

In this letter, we present a comprehensive understanding of the crossover in HBABP dynamics by studying optically trapped phoretically active microspheres, along with numerical simulations and analytical calculations. We analyzed the resultant or residual velocity ($v_{\mathrm{res}}$), which reveals the remnant propulsion speed after counteracting the restoring force field, thus providing a measure of the residual activity, together with the mean squared displacement (MSD) and power spectral density (PSD), at various settings. Our results show that the residual activity, quantified as the root mean square (rms) value of $v_{\mathrm{res}}$, is solely governed by the ratio of the characteristic timescales, persistence time ($\tau_{\mathrm{R}}$), and equilibration time in the harmonic potential ($\tau_k$). An evidently activity-dominated resultant dynamics with large residual activity at $\tau_{\mathrm{R}}/\tau_{k} \ll$ 1 becomes progressively less active with the ratio approaching 1, finally reducing to its passive equivalent, \textit{i.e.}, \textit{harmonically bound Brownian particle} (HBBP)-like motion at a radial distance where the activity is counterbalanced by restoring force as $\tau_{\mathrm{R}}/\tau_{k}$ crosses 1. The propulsion speed does not affect this dynamical crossover.

An ABP exhibits active propulsion with a constant speed $V$ along an intrinsic direction of the particle, which evolves through orientational diffusion, in addition to its passive translational diffusion, with coefficients $D_{\mathrm{R}}$ and $D_{\mathrm{T}}$, respectively. Hence, the correlation of the propulsion direction decays over a persistence time $\tau_{\mathrm{R}} = 1/D_{\mathrm{R}}$. Its activity is quantified by the P\'eclet number, $\mathrm{Pe} = V/\sqrt{D_{\mathrm{R}} D_{\mathrm{T}}}$ \cite{Bechinger2016}. When under harmonic confinement (Fig. \ref{fig:HBABP}(a, b)), the resultant HBABP dynamics is described by Langevin equations incorporating the restoring force and activity with Brownian motion (Eq. \ref{eq:LE}) \cite{tenhagenBrownianMotionSelfpropelled2011}. For the experimental realization, we used phoretically active platinum-half-coated silica (Pt-silica) Janus microspheres (diameter 2$a$ = \SI{1.76}{\um}) in an optical trap (Fig. \ref{fig:HBABP}(c, d)) \cite{Halder2026} at varying propulsion speeds $V$ and force constants $k$, which were regulated by the laser power ($P$) and \ce{H2O2} concentration ($c$) (EM). We further simulated the HBABP dynamics by numerically solving the Langevin equations (Eq. \ref{eq:LE}), where $\tau_{\mathrm{R}}$, $\tau_{k}$ ($= 6 \pi \eta a / k$, $\eta$ being the viscosity of the medium), and $V$ were varied systematically over a wide range \cite{halder2025PRE}.

\paragraph{Position distribution.} 
The analytical solution for the position distribution exhibits characteristically different features for $\tau_{\mathrm{R}} / \tau_{k} \ll 1$ and $\tau_{\mathrm{R}} / \tau_{k} \gg 1$, which therefore defines two quintessential regimes. At $\tau_{\mathrm{R}} \ll \tau_{k}$, the propulsion direction becomes random long before the system equilibrates. Consequently, $P(x)$ and $P(y)$ evolve into identical zero-mean Gaussians at $t \gg \tau_{k}$ (Eq. \ref{eq:PosDist1}), irrespective of $V$ and other initial conditions. A quadratic increase in the variance with $\mathrm{Pe}$ signifies activity-dominated dynamics in this regime \cite{halder2025PRE}. When $\tau_{\mathrm{R}} \gg \tau_{k}$, the propulsion direction persists for a longer duration, and the ABP continues to follow the initial direction (considered as $\hat{x}$) until it equilibrates, where the restoring force counterbalances its activity. Here, the activity affects only $P(x)$, which simplifies to a Gaussian with a non-zero mean at $t \gg \tau_{k}$ (Eq. \ref{eq:PosDist2}). Intriguingly, the variance of $P(x)$ remains independent of $\mathrm{Pe}$ and is equal to that of $P(y)$, which describes the corresponding passive case, \textit{i.e.}, the HBBP. However, the mean shifts linearly with $V$ as $\left\langle x \right\rangle = V \tau_{k}$ \cite{halder2025PRE}. At $t \gg \tau_{\mathrm{R}}$, the propulsion direction changes slowly, and the position distribution moves azimuthally, maintaining $\left\langle r \right\rangle = V\gamma_{\text{T}}/k$. It eventually covers an annular region and appears bimodal in both $P(x)$ and $P(y)$.

\begin{figure*}[hbt]
	\centering
	\includegraphics[width=0.9\textwidth]{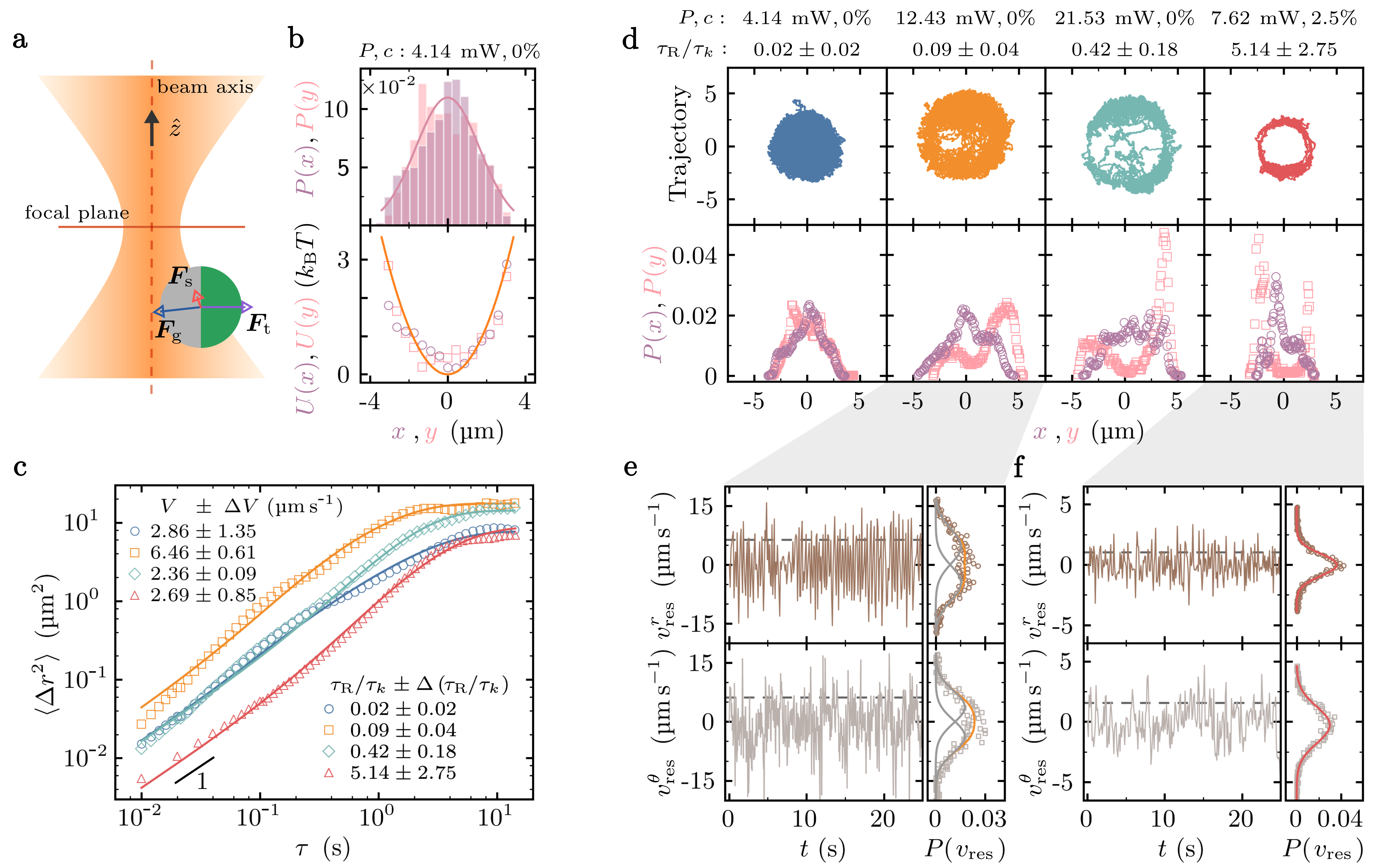}
	\caption{Experimentally observed dynamics of optically trapped active Janus colloids. (a) A Pt-silica (grey-green) Janus particle experiences scattering ($\bm{F}_{\mathrm{s}}$), gradient ($\bm{F}_{\mathrm{g}}$), and thermophoretic ($\bm{F}_{\mathrm{t}}$) forces in an optical trap (orange gradient). (b) Its stable symmetric harmonic confinement for a low laser power ($P$) is demonstrated by the Gaussian position distributions and corresponding quadratic potentials obtained by Boltzmann inversion (EM). (c) MSDs (symbols), calculated from the corresponding trajectories (the same color as in (d)), are shown for four typical cases with varied $P$ and $c$ (\ce{H2O2} concentration). Fitting the MSDs to Eq. \ref{eq:MSDGen} (solid lines) provides the respective $V$, $\tau_{\mathrm{R}}$, and $\tau_{k}$. (d) Trajectories and corresponding position distributions are exhibited in the order of increasing $\tau_{\mathrm{R}}/\tau_{k}$. (e, f) Short segments of the time series and probability distributions (symbols) of $v_{\mathrm{res}}^{r}$ and $v_{\mathrm{res}}^{\theta}$ (Eq. \ref{eq:v_res}) are shown for the two characteristic cases. The velocity distributions for (e) $\tau_{\mathrm{R}}/\tau_{k} <$ 1 fit well with the sum of two symmetrically positioned Gaussians (solid lines), whereas those for (f) $\tau_{\mathrm{R}}/\tau_{k} >$  1 show an excellent fit to zero-mean Gaussians (solid lines). The rms values of the velocity components are marked with dashed horizontal lines.
		\label{fig:Expt}}
\end{figure*}

The experimentally observed dynamics of an optically trapped Janus particle (Fig. \ref{fig:Expt}(d)) exhibited the predicted crossover governed by $\tau_{\mathrm{R}}/\tau_{k}$. For $\tau_{\mathrm{R}}/\tau_{k} \ll$ 1 (0.02 and 0.09), the ABP spontaneously crossed the center (Supp. Video 1) and the spread of the space-filling bound trajectories increased with $V$. As $\tau_{\mathrm{R}}/\tau_{k}$ increased, the trajectories progressively avoided the central region and eventually became annularly confined at $\tau_{\mathrm{R}}/\tau_{k}$ = 5.14 (Supp. Video 2). Corresponding position distributions, $P(x)$ and $P(y)$, even before reaching steady-state, show signatures of the crossover from zero-mean Gaussian (Eq. \ref{eq:PosDist1}) to a bimodal distribution (Eq. \ref{eq:PosDist2}) with increasing $\tau_{\mathrm{R}}/\tau_{k}$, irrespective of $V$.

\begin{figure*}[htbp]
	\centering
	\includegraphics[width=0.85\textwidth]{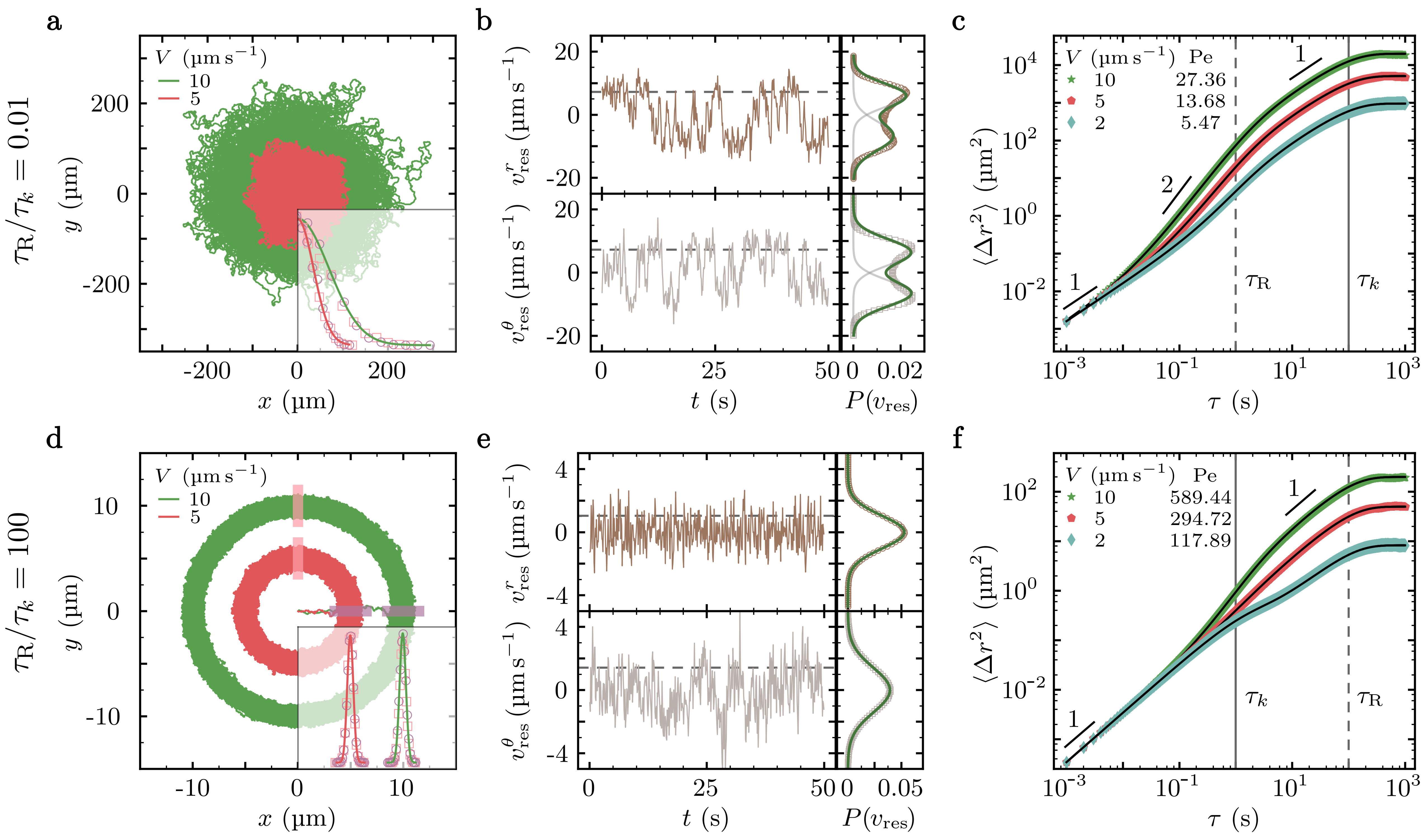}
	\caption{Numerically simulated dynamics of HBABP with $\tau_{\mathrm{R}}/\tau_{k} \ll$ 1 (top row) and $\tau_{\mathrm{R}}/\tau_{k} \gg$ 1 (bottom row). (a, d) Trajectories and corresponding position distributions $P(x)$ (circles) and $P(y)$ (squares) from the positive sides (right bottom inset) are shown for two propulsion speeds ($V$). In (d), $P(x)$ and $P(y)$ are calculated from narrow regions (highlighted) around the $x$-and $y$ axes, respectively. The position distributions show excellent fitting (solid lines) with (a) Eq. \ref{eq:PosDist1} and (d) Eq. \ref{eq:PosDist2}. (b, e) Short segments of the time series and probability distributions (symbols) of $v_{\mathrm{res}}^{r}$ and $v_{\mathrm{res}}^{\theta}$ (Eq. \ref{eq:v_res}) are shown for $V$ = \SI{10}{\um /s}. The probability distributions in (b) fit well to the sum of the two symmetrically positioned Gaussians (solid lines), whereas those in (e) exhibit an excellent fit with single zero-mean Gaussians (solid lines). The horizontal dashed lines represent the respective rms values. (c, f) MSDs (symbols) for three $V$ values are shown with fits to Eq. \ref{eq:MSDGen} (solid lines). The corresponding $\tau_{\mathrm{R}}$ and $\tau_{k}$ values are marked by dashed and solid vertical lines, respectively.
		\label{fig:Sim}}
\end{figure*}

The simulated steady-state position distributions at higher $V$ make the analytical prediction of the $\tau_{\mathrm{R}}/\tau_{k}$-dependent crossover more apparent (Fig. \ref{fig:Sim}(a, d)). For $\tau_{\mathrm{R}}/\tau_{k}$ = 0.01, the variances of $P(x)$ and $P(y)$, which are zero-mean Gaussians, increase with $V$, validating Eq. \ref{eq:PosDist1} (Fig. \ref{fig:Sim}(a), \ref{fig:kEff}(a)). In contrast, the peaks of $P(x)$ and $P(y)$ shift outwards with increasing $V$, while they fit well to Gaussians with unchanged variance, following Eq. \ref{eq:PosDist2}, at $\tau_{\mathrm{R}}/\tau_{k}$ = 100 (Fig. \ref{fig:Sim}(d), \ref{fig:kEff}(d)). The corresponding dynamics (Supp. Videos 3 and 4) are manifested in the space-filling bound and annularly confined long-time trajectories (Fig. \ref{fig:Sim}(a, d)), respectively.

\paragraph{Residual velocity.} 
To further examine the signature of activity in the HBABP dynamics, we analyzed the residual or resultant velocity $v_{\mathrm{res}}$ in the restoring force field, as described by Eq. \ref{eq:LE}. Exploiting circular symmetry, we computed its radial ($v_{\mathrm{res}}^{r}$) and azimuthal ($v_{\mathrm{res}}^{\theta}$) components (EM), which denote:
\begin{equation}
	v_{\mathrm{res}}^{r} = V^{r} - r / \tau_k + v_{\mathrm{B}}^{r} \qquad  \text{and} \qquad v_{\mathrm{res}}^{\theta} = V^{\theta} + v_{\mathrm{B}}^{\theta}.
	\label{eq:v_res}
\end{equation}

For $\tau_{\mathrm{R}}/\tau_{k} \ll$ 1, both components of $v_{\mathrm{res}}$ have significantly higher rms and maximum values than those in cases with $\tau_{\mathrm{R}}/\tau_{k} \gg$ 1 (Fig. \ref{fig:Expt}(e, f) and \ref{fig:Sim}(b, e)). The probability distributions at $\tau_{\mathrm{R}}/\tau_{k} \ll$ 1 are flat-topped Gaussians (Fig. \ref{fig:Expt}(e)) or bimodal (Fig. \ref{fig:Sim}(b)), consisting of two symmetrically placed Gaussians, with peaks nearly at the corresponding rms value. Moreover, they follow $\left\langle \left( v_{\mathrm{res}}^{r} \right) ^2\right\rangle + \left\langle \left( v_{\mathrm{res}}^{\theta} \right) ^2\right\rangle \approx V^2 $, indicating that the ABP retains mostly the same propulsion speed even under harmonic confinement. In this regime, the propulsion direction changes rapidly; consequently, a fast-varying $V^{r} (t)$ is rarely balanced by the instantaneous position-dependent term $r (t) / \tau_{k}$, whereas $V^{\theta}$ remains unimpeded by the radial restoring force field. At $\tau_{\mathrm{R}}/\tau_{k} \gg$ 1, the probability distributions of both $v_{\mathrm{res}}^{r}$ and $v_{\mathrm{res}}^{\theta}$ are zero-mean Gaussian with significantly smaller rms values, resembling the HBBP dynamics (Fig. \ref{fig:Expt}(f), \ref{fig:Sim}(e)). Here, with a slowly changing propulsion direction, the dominating component $V^{r}$ is continually counterbalanced by $r / \tau_{k}$ as the ABP remains radially confined around $r = V^{r} \tau_{k}$, while a slowly developing $V^{\theta}$ leads the ABP to an azimuthal position ($\theta$) where it vanishes and thus remains systematically insignificant.

Intriguingly, the probability distributions of the residual velocity components in both regimes apparently contradict the steady-state position distribution \cite{halder2025PRE}. However, they both signify the same crossover, which is from activity-dominated dynamics at $\tau_{\mathrm{R}}/\tau_{k} \ll$ 1 to activity-depleted HBBP dynamics at $\tau_{\mathrm{R}}/\tau_{k} \gg$ 1, where the activity is neutralized by the restoring force at a radially displaced position.

\paragraph{Mean square displacement.}
The trend of the MSDs also manifests the characteristics of the dynamics at varied timescales. When $\tau_{\mathrm{R}}/\tau_{k} \ll$ 1, the MSD (Eq. \ref{eq:MSD1}) at intermediate time-lags, up to $\tau_{\mathrm{R}}$, is dominated by a term that varies as $V^{2} \tau^{2}$, verifying the activity-driven resultant dynamics in this regime. At $\tau_{\mathrm{R}}/\tau_{k} \gg$ 1, the MSD (Eq. \ref{eq:MSD2}) is the sum of two terms that increase linearly before reaching plateaus at $\tau > \tau_{k}$ and $\tau > \tau_{\mathrm{R}}$, respectively. This indicates the absence of steady activity in the resultant dynamics at $\tau_{\mathrm{R}}/\tau_{k} \gg$ 1.

The MSDs of the active Janus particles in an optical trap fit well with the analytical prediction, \textit{i.e.}, Eq. \ref{eq:MSDGen} (Fig. \ref{fig:Expt}(c)). Despite having similar $V$, the MSDs with $\tau_{\mathrm{R}}/\tau_{k}$ = 0.02 and 0.42 ($<$ 1) are significantly higher than that with $\tau_{\mathrm{R}}/\tau_{k}$ = 5.14 and manifest activity-dominated dynamics. The presence of activity in the resultant dynamics at $\tau_{\mathrm{R}}/\tau_{k}$ = 0.02 and 0.09 is further corroborated by a 4 to 6 times increase in MSD as $V$ becomes $\sim$ 2.5 times (from \SI{2.86}{\um /s} to \SI{6.46}{\um /s}).

The MSDs from the simulated dynamics are also in excellent agreement with Eq. \ref{eq:MSDGen} (Fig. \ref{fig:Sim}(c, f)). At $\tau_{\mathrm{R}}/\tau_{k}$ = 0.01 (Fig. \ref{fig:Sim}(c)), the clear emergence of $\tau^2$ dependence as $\tau$ approaches $\tau_{\mathrm{R}}$ validates the activity-dominated dynamics. For $\tau_{\mathrm{R}}/\tau_{k}$ = 100 (Fig. \ref{eq:MSDGen}(f)), the MSDs mostly increase linearly before saturation and remain ~$\sim$ 100 times smaller than those with the same $V$ values in the former case, despite having a 20 times higher $\mathrm{Pe}$. While the MSD for $V$ = \SI{2}{\um /s} shows two plateaus as predicted by Eq. \ref{eq:MSD2}, the plateau at shorter $\tau$ remains suppressed by the subsequent rise for larger $V$.

\begin{figure}[htb]
	\centering
	\includegraphics[width=0.95\linewidth]{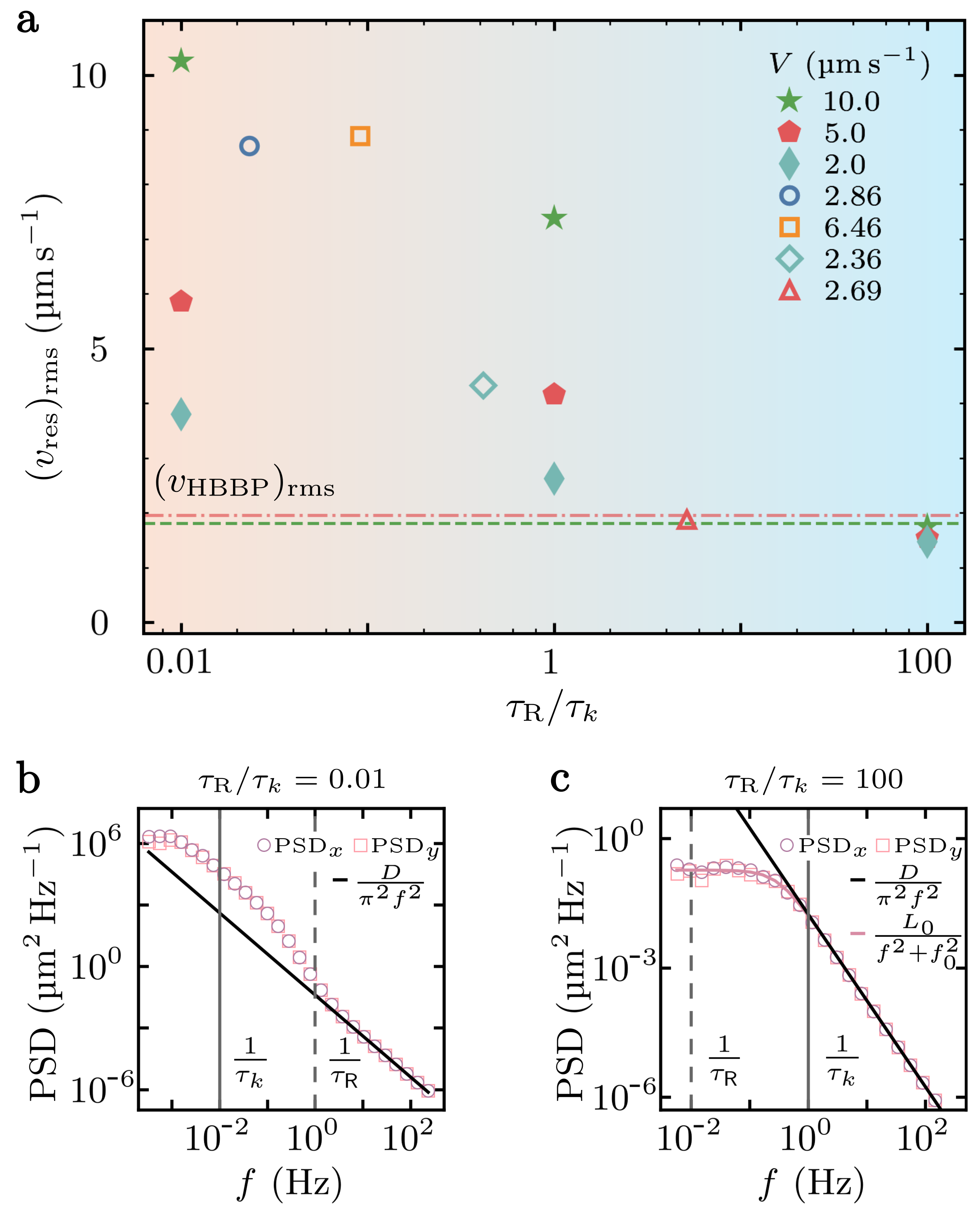}
	\caption{Variation in residual activity, quantified as $\left( v_{\mathrm{res}} \right)_{\mathrm{rms}}$, and PSDs with $\tau_{\mathrm{R}}/\tau_{k}$. (a) $\left( v_{\mathrm{res}} \right)_{\mathrm{rms}}$ is calculated and shown for varied $\tau_{\mathrm{R}}/\tau_{k}$ values from experiments (open symbols) and simulations (filled symbols). The corresponding $V$ values are presented in the inset. For cases with $\tau_{\mathrm{R}}/\tau_{k} >$ 1, the theoretical values of $\left( v_\mathrm{HBBP} \right) _\mathrm{rms}$, \textit{i.e.}, the respective $\left( v_{\mathrm{res}} \right)_{\mathrm{rms}}$ with $V$ = 0, are indicated by the color-coded dashed horizontal lines; the $\left( v_\mathrm{HBBP} \right) _\mathrm{rms}$ values (EM) for the three simulated trajectories with $\tau_{\mathrm{R}}/\tau_{k} =$ 100 and varied $V$ are the same and are marked by the green dashed line. The background color gradient implies a crossover in the residual activity with $\tau_{\mathrm{R}}/\tau_{k}$. The PSDs from the simulated dynamics with $V$ = \SI{10}{\um /s} are compared with the FDT predictions (EM) (black solid lines) for (b) $\tau_{\mathrm{R}}/\tau_{k} \ll$ 1 and (c) $\tau_{\mathrm{R}}/\tau_{k} \gg$ 1. In (c), $\mathrm{PSD}_x$ and $\mathrm{PSD}_y$ are obtained from the trajectory segments around the $x$- and $y$-axes (highlighted in \ref{fig:Sim}(d)), respectively, and the Lorentzian fitting is shown by a magenta solid line.
	\label{fig:ResAct}}
\end{figure}

\paragraph{Residual activity.}
To illustrate the crossover from activity-dominated to activity-depleted HBBP-like resultant dynamics as $\tau_{\mathrm{R}}$ becomes longer than $\tau_{k}$, we quantify the residual activity as $\left( v_{\mathrm{res}} \right)_{\mathrm{rms}}$, \textit{i.e.}, the rms value of $v_{\mathrm{res}}$ (Eq. \ref{eq:v_res}). It decreases with increasing values of t$\tau_{\mathrm{R}}/\tau_{k}$, undergoing a crossover at $\tau_{\mathrm{R}}/\tau_{k}$ = 1, irrespective of $V$ and $\mathrm{Pe}$ (Fig. \ref{fig:ResAct}(a)). At $\tau_{\mathrm{R}}/\tau_{k} \ll$ 1, $\left( v_{\mathrm{res}} \right)_{\mathrm{rms}}$ exceeds $V$ by an additional contribution from $v_{\mathrm{HBBP}}$. Maintaining a decreasing trend, $v_{\mathrm{res}}$ retains a portion of the propulsion speed at $\tau_{\mathrm{R}}/\tau_{k}$ = 1 (EM). At $\tau_{\mathrm{R}}/\tau_{k} >$ 1, all $\left( v_{\mathrm{res}} \right)_{\mathrm{rms}}$ values converge to the corresponding $\left( v_{\mathrm{HBBP}} \right)_{\mathrm{rms}}$ (EM), verifying the absence of residual activity as the propulsion is neutralized by the restoring force at $r = V^{r} \tau_{k}$.

\paragraph{Effective confinement and PSD.}
We analyzed the effective harmonic confinement (EM) and PSD of the resultant dynamics in the two characteristic regimes to further verify our findings. At $\tau_{\mathrm{R}}/\tau_{k} \ll$ 1, the effective force constant $k_{\mathrm{eff}}$ decreases with increasing $V$ (Fig. \ref{fig:kEff}(c)), and the PSDs deviate from fluctuation dissipation theorem (FDT) prediction (EM) at lower frequencies, where propulsion dominates the Brownian dynamics (Fig. \ref{fig:ResAct}(b)), indicating activity-dominated resultant dynamics. For $\tau_{\mathrm{R}}/\tau_{k} \gg$ 1, $k_{\mathrm{eff}}$ remains unchanged, and the radial distance of the center of the confinement $r_{\mathrm{c}}$ increases linearly with $V$ (Fig. \ref{fig:kEff}(f)). Furthermore, the PSDs fit perfectly to Lorentzian (EM) and validate the FDT beyond the corner frequency (Fig. \ref{fig:ResAct}(c)), confirming HBBP-like resultant dynamics devoid of activity, which is depleted to displace the harmonic well radially outward.

\paragraph{Conclusions and discussions.}
In this \textit{Letter}, we present detailed analyses of the position distribution, resultant velocity and residual activity, MSD, effective confinement descriptions, and PSD of the resultant dynamics of an HBABP employing experiments, numerical simulations, and analytical calculations. Our findings demonstrate a crossover from activity-dominated bound dynamics at $\tau_{\mathrm{R}}/\tau_{k} <$ 1 to activity-depleted HBBP-like motion in annular confinement for $\tau_{\mathrm{R}}/\tau_{k} >$ 1. We further show that the propulsion speed does not affect this crossover, which is solely governed by the interplay between the characteristic timescales $\tau_{\mathrm{R}}$ and $\tau_{k}$.

Our results imply that, in general, active systems exhibit their inherent active dynamics within a bound when the persistence time is shorter than the timescale associated with confinement. In contrast, intriguing dynamical features emerge as their long-persistent propulsion drives them directly to the boundary, where their activity competes with confinement and is depleted. The crossover from one regime to the other is controlled by the ratio of the timescales associated with the persistence of motion and confinement. Thus, our conclusions have far-reaching implications for understanding the dynamical phenomena in various confined active matters, such as accumulation near the boundary \cite{Berke2008, elgetiWallAccumulationSelfpropelled2013, leeActiveParticlesConfinement2013}, ordering \cite{bauerleFormationStableResponsive2020, lavergneGroupFormationCohesion2019, hennesSelfInducedPolarOrder2014}, and active glasses under crowding \cite{Ni2013, Berthier2013, Janssen2019}. Furthermore, our study provides a reliable quantification of the manifestation of activity and, thereby, the deviation from equilibrium \cite{Fodor2016a, Dabelow2021}, in addition to offering a viable approach for describing the mechanical pressure of active systems under confinement \cite{Yan2015}.

Studying and comparing the residual activities in run-and-tumble and chiral ABP motions in harmonic and anharmonic confinements \cite{solonActiveBrownianParticles2015, Caprini2023} remain important future directions. Additionally, confining ABPs to dynamically varying potentials, \textit{e.g.}, time-dependent or slow-moving optical traps \cite{Khan2014,Khan2014a, Liu2018, Halder2024, Halder2026a}, provides an excellent means to study their interactions with complex environments.

\paragraph{Acknowledgment.}
The authors acknowledge the Science and Engineering Research Board (SERB), Govt. of India, for supporting this work through a Core Research Grant (CRG/2020/002723), and the PARAM Sanganak computing facility at the Computer Center, IIT Kanpur, for the numerical simulations. MK thanks Abhik Basu, Ambarish Ghosh, and Sriram Ramaswamy for fruitful discussions and critical reading of the manuscript.






\bibliography{ABPinHW_references.bib}


\section{End Matter}

\paragraph{Free ABP dynamics.} 
The 2D MSD of a free ABP is given by \cite{howseSelfMotileColloidalParticles2007, Bechinger2016, basuActiveBrownianMotion2018}:
\begin{equation}
\left\langle \Delta r^2 (\tau) \right\rangle  = \left(4D_{\mathrm{T}} + 2 V^2 \tau_{\mathrm{R}} \right)\tau + 2 V^2 \tau_{\mathrm{R}}^2 \left( e^{- \tau / \tau_{\mathrm{R}}} - 1 \right),
\label{eq:ABP-MSD}
\end{equation}
and the relative contribution of activity to the dynamics is defined by the P\'eclet number, $\mathrm{Pe} = V/\sqrt{D_{\mathrm{R}} D_{\mathrm{T}}}$ \cite{Bechinger2016}.

\begin{figure}[htb]
	\centering
	\includegraphics[width=0.8\linewidth]{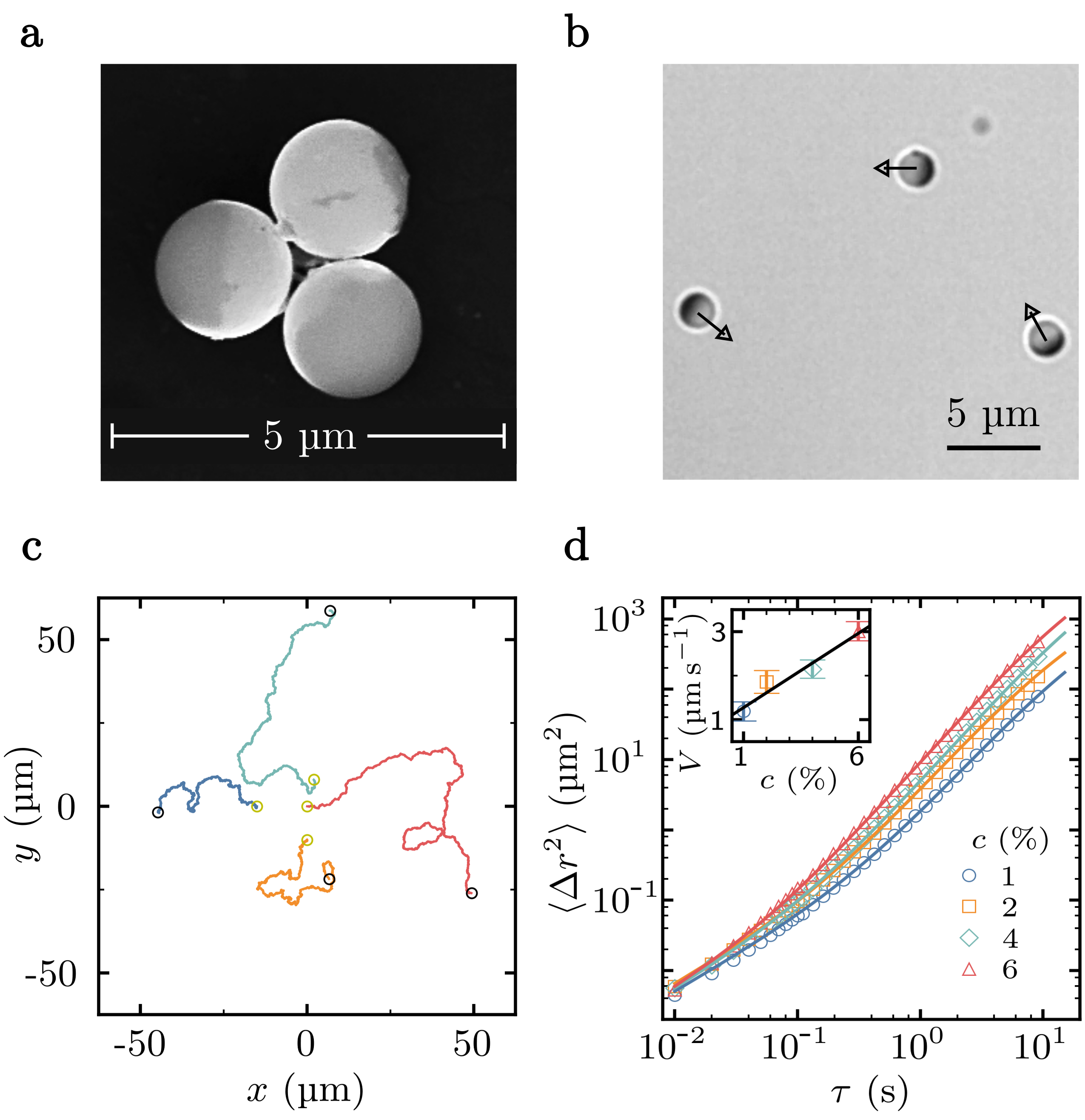}
	\caption{Diffusiophoretic active dynamics of Pt-silica Janus colloids. (a) FESEM image of the half-Pt-coated (appears brighter) silica particles and (b) their bright-field micrograph, where the Pt coating appears darker, are exhibited. The arrows denote the direction of propulsion. (c) Four typical color-coded trajectories of free Janus colloids at various \ce{H2O2} concentrations are shown with (d) corresponding averaged MSDs (symbols). The MSDs are fitted to Eq. \ref{eq:ABP-MSD} (solid lines) to obtain $V$, which exhibits a linear increase with $c$ (inset).} 
	\label{fig:FreeABP}
\end{figure}

ABPs were experimentally realized using Pt-silica Janus colloids (Fig. \ref{fig:FreeABP}(a, b)), which experience active propulsion owing to self-diffusiophoresis in an \ce{H2O2} solution \cite{howseSelfMotileColloidalParticles2007} and self-thermophoresis under laser exposure \cite{jiangActiveMotionJanus2010,Halder2026}. The synthesis of Pt-silica Janus colloids and calibration of their thermophoretic activity have been discussed in our previous study \cite{Halder2026}. The diffusiophoretic activity of these Janus microspheres was characterized at various concentrations ($c$\% in vol/vol) of \ce{H2O2} by recording the trajectories in the absence of a laser trap and fitting the averaged MSDs using Eq. \ref{eq:ABP-MSD} (Fig. \ref{fig:FreeABP}(c, d)).

\paragraph{Theoretical description of HBABP.} 
The Langevin equations describing the translational ($x(t)$, $y(t)$) and orientational ($\phi(t)$) dynamics of an ABP in a harmonic well with force constant $k$ (Fig. \ref{fig:HBABP}) can be written as \cite{tenhagenBrownianMotionSelfpropelled2011}:
\begin{align}
		\dot{x} &= v_{\mathrm{B}}^x - x / \tau_k + V \cos{\phi }, \quad  \dot{y} = v_{\mathrm{B}}^y - y / \tau_k + V \sin{\phi},  \; \text{and} \nonumber \\
		\dot{\phi} &=  v_{\mathrm{B}}^{\phi},   	\label{eq:LE}
\end{align}
where $v^i_{\mathrm{B}}$, $\tau_k = \gamma_{\text{T}} / k$, and $\gamma_{\text{T}}$ are the Brownian velocities along $\hat{i}$ ($i = x, y, \phi$), equilibration time, and translational Stokes drag coefficient, respectively. 

Closed-form solutions for the position distributions can be derived for $\tau_{\mathrm{R}} / \tau_{k} \ll 1$ and $\tau_{\mathrm{R}} / \tau_{k} \gg 1$ \cite{halder2025PRE}. At $\tau_{\mathrm{R}} \ll \tau_{k}$, $P(x)$ and $P(y)$ are given by the same form,
\begin{equation}
	P(x) = \frac{1}{\sqrt{\frac{2 \pi k_{\mathrm{B}} T}{k} \left( 1 + \frac{\mathrm{Pe}^2}{2} \right) }} \exp\left[ {\frac{- x^2}{\frac{2 k_{\mathrm{B}} T}{k} \left( 1 + \frac{\mathrm{Pe}^2}{2} \right) }}\right].
	\label{eq:PosDist1}
\end{equation}
For $\tau_{\mathrm{R}} \gg \tau_{k}$, the position distribution along the initial propulsion direction, say $\hat{x}$, becomes (at $\tau_{\mathrm{R}} \gg t \gg \tau_{k}$)
\begin{equation}
	P(x) = \frac{1}{\sqrt{\frac{2 \pi k_{\mathrm{B}} T}{k} }} \exp\left[ \frac{- \left( x - V \tau_{k}\right) ^2}{2 k_{\mathrm{B}} T / k} \right].
	\label{eq:PosDist2}
\end{equation}
$P(y)$ takes a similar form, but with $V$ = 0.

\begin{figure*}[htb]
	\centering
	\includegraphics[width=0.7\textwidth]{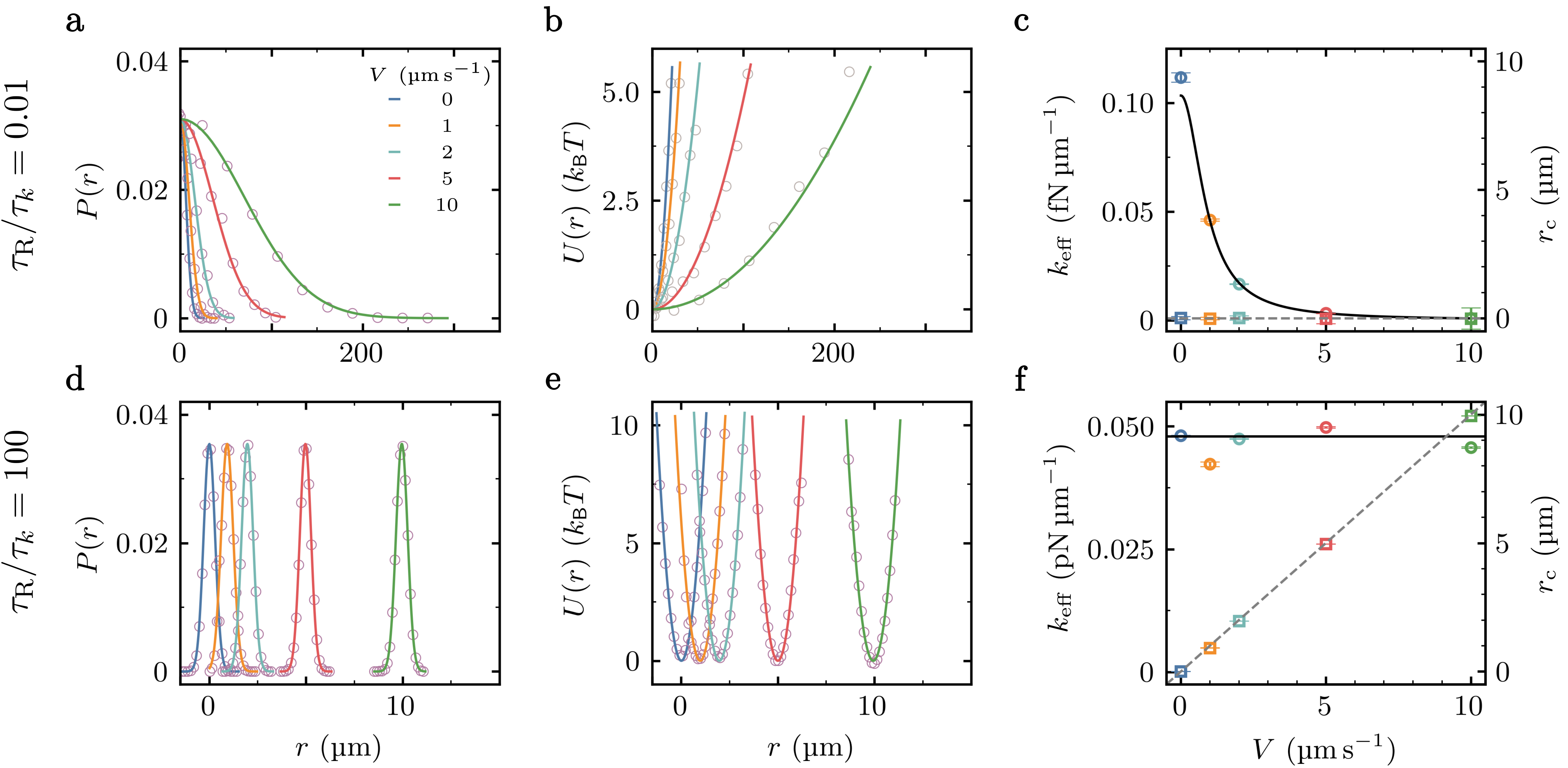}
	\caption{Variation in the steady-state position distributions and corresponding effective harmonic confinement with $V$ at $\tau_{\mathrm{R}} / \tau_k \ll$ 1 (top row) and $\tau_{\mathrm{R}} / \tau_k \gg$ 1 (bottom row). (a, d) $P (r)$ from simulated HBABP dynamics, (b, e) effective quadratic potentials $U (r)$, obtained from $P (r)$ using Boltzmann inversion, and (c, f) the effective harmonic confinement parameters $k_{\mathrm{eff}}$ and $r_{\mathrm{c}}$ are shown for varied $V$. The color-coded open symbols represent the data, and the solid lines of the same color exhibit fit to (a, d) Eq. \ref{eq:PosDist1}, \ref{eq:PosDist2}, and (b, e) $\frac{1}{2} k_{\mathrm{eff}}(r - r_{\mathrm{c}})^2 $. (c, f) The variations in $k_{\mathrm{eff}}$ and $r_{\mathrm{c}}$, obtained from the fitting, show excellent agreement with the theoretical predictions (Eq. \ref{eq:k_eff1}, \ref{eq:k_eff2}) as shown by the black solid and gray dashed lines, respectively.} 
	\label{fig:kEff}
\end{figure*}

The 2D MSD is analytically derived as \cite{halder2025PRE},
\begin{multline}
	\left\langle \Delta r^{2} (\tau) \right\rangle = \frac{4 k_{\mathrm{B}} T}{k}\left[ 1 - e^{- \tau / \tau_{k}} \right]  \\ 
	+ \frac{2 V^2 \tau_{k}^2 \tau_{\mathrm{R}}}{\tau_{\mathrm{R}} + \tau_{k}}  \left[ 1 - \frac{\tau_{\mathrm{R}} e^{- \tau / \tau_{\mathrm{R}}} - \tau_{k} e^{- \tau / \tau_{k}}}{\tau_{\mathrm{R}} - \tau_{k}} \right].
	\label{eq:MSDGen}
\end{multline}
At $\tau_{\mathrm{R}}/\tau_{k} \ll$ 1, the RHS of Eq. \ref{eq:MSDGen} becomes
\begin{equation}
 \left( \frac{4 k_{\mathrm{B}} T}{k} + 2 V^2 \tau_{k} \tau_{\mathrm{R}}\right) \left[ 1 - e^{- \frac{\tau}{\tau_{k}}}\right] + 2 V^2 \tau_{\mathrm{R}}^2 e^{- \frac{\tau}{\tau_{\mathrm{R}}}}.
	\label{eq:MSD1}
\end{equation}
For $\tau_{\mathrm{R}}/\tau_{k} \gg$ 1, the MSD takes the form
\begin{equation}
	\left\langle \Delta r^{2} \right\rangle = \frac{4 k_{\mathrm{B}} T}{k}\left[ 1 - e^{- \tau / \tau_{k}} \right]  + 2 V^{2} \tau_{k}^2 \left[ 1 - e^{- \tau / \tau_{\mathrm{R}}} \right].
	\label{eq:MSD2}
\end{equation}

\paragraph{Experimental realization of HBABP.} 
The optical trapping of phoretically active Janus particles has been discussed in our previous study \cite{Halder2026}. We used a linearly polarized \SI{1064}{\nm} Nd:YAG CW laser with a Gaussian intensity profile to set up an optical trap that provided a harmonic confinement. The disparate optical properties of the two sides of a Pt-silica Janus microsphere give rise to a local temperature gradient and hence a thermophoretic force $\bm{F}_{\mathrm{t}}$, in addition to the scattering force $\bm{F}_{\mathrm{s}}$ and gradient force $\bm{F}_{\mathrm{g}}$ (Fig. \ref{fig:Expt}(a)). The relative strength and direction of these forces vary with both the instantaneous position and orientation of the particle; however, the net force, averaged over the possible position-orientations of the Janus particle, provides a stable symmetric harmonic confinement \cite{Halder2026}. This is validated by the Gaussian position distributions; therefore, quadratic potentials along both $x$ and $y$ (Fig. \ref{fig:Expt}(b)), which follow the prediction for effective confinement (Eq. \ref{eq:k_eff1}) in that setting ($\tau_{\mathrm{R}}/\tau_{k} =$ 0.02). The effective orientational diffusion, and hence $\tau_{\mathrm{R}}$ is also regulated by the asymmetric optical forces \cite{jiangActiveMotionJanus2010, merktCappedColloidsLightmills2006, nedevOpticallyControlledMicroscale2015, zongOpticallyDrivenBistable2015, Liu2015, brontecirizaOpticallyDrivenJanus2023}.

In the experimental realization, the variations in activity and confinement are coupled. We varied both $P$ and $c$ to achieve different settings in the parameter space. A lower $P$ resulted in a longer $\tau_{k}$ and a lower thermophoretic propulsion speed $V$, where $c$ was varied to increase $V$ with supplementary diffusiophoretic activity. In contrast, $P$ was increased to enhance $V$ and shorten $\tau_{k}$ and $\tau_{\mathrm{R}}$.

The ABP and HBABP dynamics were recorded at \SI{500}{\Hz}, and $x(t)$, $y(t)$, and $\phi(t)$ were obtained from the image sequences using the TrackMate ImageJ plugin \cite{ershovTrackMate7Integrating2022}. While $x(t)$ and $y(t)$ were analyzed to obtain the position and velocity distributions and MSDs, $\phi(t)$ provided an estimation of $\tau_{\mathrm{R}}$. We determined the values of $V$, $\tau_{\mathrm{R}}$, and $\tau_{k}$ by fitting the experimental MSDs with Eq. \ref{eq:MSDGen} using the weighted least squares (WLS) regression method \cite{baileyFittingActiveBrownian2022}.


\paragraph{Calculation of residual velocity.} 
To calculate the instantaneous residual velocity components, $v_{\mathrm{res}}^r = \dot{r}$ and $v_{\mathrm{res}}^{\theta} = r \dot{\theta}$ from the trajectories, we retained all relevant and crucial dynamical information in $v_{\mathrm{res}}$ but eliminated redundant high-frequency fluctuations by applying appropriate averaging over $r(t)$ and $\theta(t)$.

\paragraph{Effective harmonic confinement.} 
Comparing the steady-state position distributions (Eq. \ref{eq:PosDist1}, \ref{eq:PosDist2}), and MSDs (Eq. \ref{eq:MSD1}, \ref{eq:MSD2}) to those of an HBBP, we define the effective harmonic confinement in the two regimes with the corresponding force constant $k_{\mathrm{eff}}$ and the radial position of the center $r_{\mathrm{c}}$ as \cite{halder2025PRE}:
\begin{align}
		k_{\mathrm{eff}} & = \frac{k}{ 1 + \mathrm{Pe}^2 / 2},  &  r_{\mathrm{c}} & = 0 &  \text{for}    \ \tau_{\mathrm{R}} \ll \tau_{k}, 
		\label{eq:k_eff1}\\ 
		k_{\mathrm{eff}}  &  = k,  & r_{\mathrm{c}} & = V \tau_k & \text{for} \ \tau_{\mathrm{R}} \gg \tau_{k}.
		\label{eq:k_eff2}
\end{align}
We obtain $k_{\mathrm{eff}}$ and $r_{\mathrm{c}}$ from the simulated HBABP position distributions at various $V$ and the variations show excellent agreement with  Eq. \ref{eq:k_eff1} and \ref{eq:k_eff2} (Fig. \ref{fig:kEff}).

\paragraph{FDT prediction and $(v_\mathrm{HBBP})_\mathrm{rms}$.}
Conforming to FDT, the PSD of a HBBP is given by, $\mathrm{PSD}_x = \mathrm{PSD}_y = \frac{k_{\mathrm{B}}T}{\pi^2 \gamma_{\mathrm{T}} \left( f^2 + f_{0}^2 \right)} = \frac{L_{0}^2}{f^2 + f_{0}^2} $,which is a Lorentzian with corner-frequency $f_{0} = \left( 2 \pi \tau_k\right) ^{-1}$. Furthermore, the rms value of $v_{\mathrm{HBBP}}$ (\textit{i.e.} for $V$ = 0) in 2D is given by $(v_\mathrm{HBBP})_\mathrm{rms} = \sqrt{(4D_{\text{T}} / \Delta t) - (2D_{\text{T}} / \tau_{k})}$, where $\Delta t$ = \SI{0.1}{\s}. For a free Brownian particle, $f_{0}$ = 0; therefore, $\mathrm{PSD}_x = \mathrm{PSD}_y = D_{\mathrm{T}} / \pi^2 f^2$ \cite{halder2025PRE}.

\begin{figure}[htbp]
	\centering
	\includegraphics[width=0.85\linewidth]{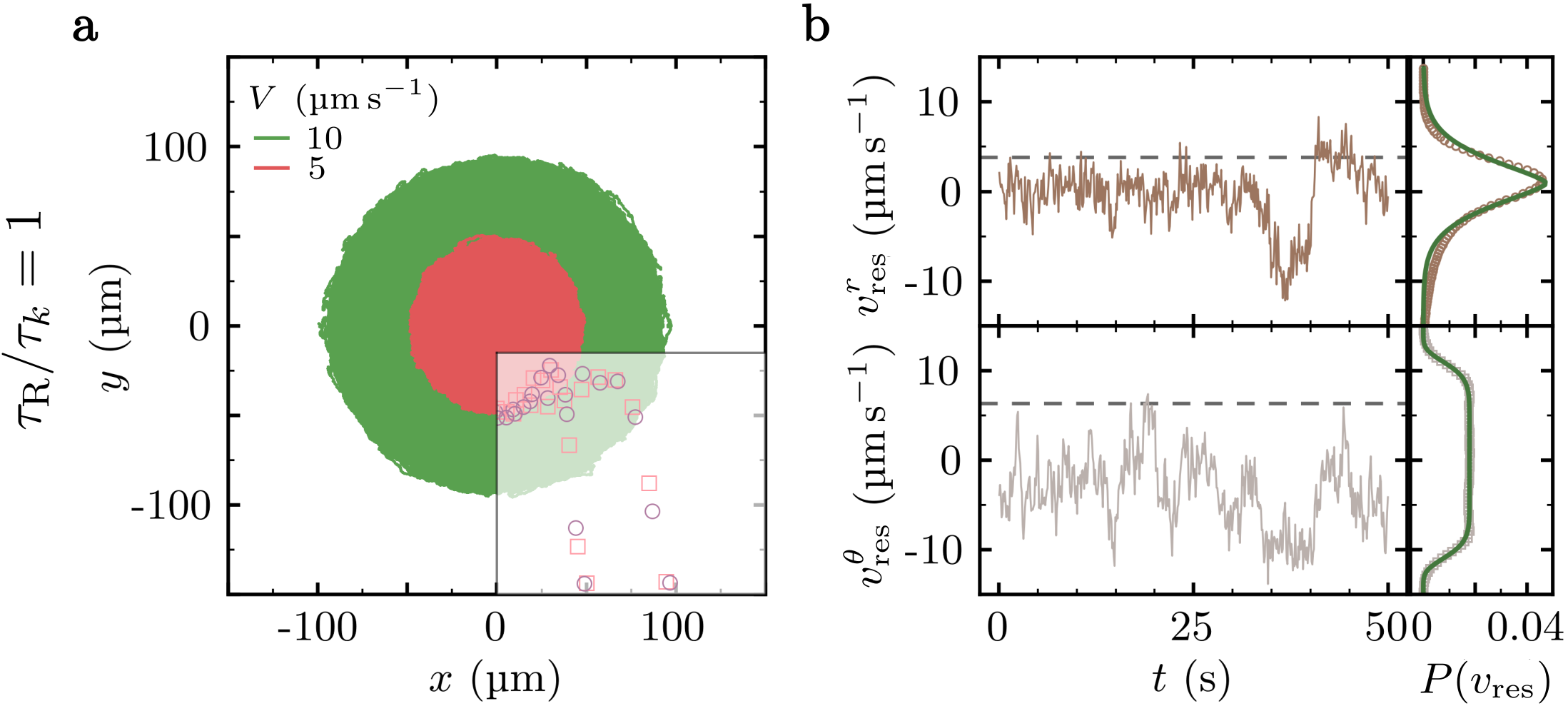}
	\caption{Numerically simulated HBABP dynamics, similar to those in Fig. \ref{fig:Sim}, but with $\tau_{\mathrm{R}} = \tau_{k}$ = \SI{10}{\s}. (a) Trajectories and corresponding position distributions for two values of $V$. (b) Sort segments of the time series and probability distributions of the $v_{\mathrm{res}}$ components for $V$ = \SI{10}{\um /s}. The fitting of $P(v_{\mathrm{res}}^{r})$ and $P(v_{\mathrm{res}}^{\theta})$ with peaked and flat-topped Gaussians, respectively, are shown by solid lines. } 
	\label{fig:equal}
\end{figure}

\paragraph{The intermediate case.}
Simulated HBABP dynamics with $\tau_{\mathrm{R}}/\tau_{k} =$ 1 exhibit center-avoiding but space-filling trajectories with bimodal position distributions (Fig. \ref{fig:equal}(a)). $v_{\mathrm{res}}^r$ has a relatively smaller rms value with a super-Gaussian distribution, whereas $v_{\mathrm{res}}^{\theta}$ shows a flat-topped Gaussian distribution with a higher rms value (Fig. \ref{fig:equal}(b)); both deviate from the corresponding HBBP behavior, validating the presence of weak activity \cite{halder2025PRE}.

\end{document}


\title{Supplemental Material for \\``Interplay between timescales governs the residual activity of a harmonically bound active Brownian particle''}


\author{Sanatan Halder \orcidlink{0009-0002-2457-0449}}
\email[Contact author: ]{sanatanh@iitk.ac.in}
\affiliation{Department of Physics, Indian Institute of Technology Kanpur, Kanpur - 208016, India}

\author{Manas Khan \orcidlink{0000-0001-6446-3205}}
\email[Contact author: ]{mkhan@iitk.ac.in}
\affiliation{Department of Physics, Indian Institute of Technology Kanpur, Kanpur - 208016, India}


\maketitle

\onecolumngrid

\section*{Supplementary Videos}

\noindent
\href{https://drive.google.com/file/d/1B5QWNyLm38tlgPe1so7F3_dVuxH8g7lA/view?usp=drive_link}{\textbf{Supp. Video 1}}

\vspace*{0.1cm}
\noindent
Thermophoretic active dynamics of a Pt-silica Janus colloid (diameter $2a$ = \SI {1.76}{\micro \meter}) in an optical trap at a laser power $P$ = \SI{4.14}{\mW} (Fig. 2(d)). The values of the relevant parameters in this setting are given by the propulsion speed $V$ = \SI {2.86}{\micro \meter /s}, and the ratio of the persistence time ($\tau_{\mathrm{R}}$) and equilibration time ($\tau_k$), $\tau_{\mathrm{R}}/\tau_k$ = 0.02. The dynamics was observed under a microscope with a 60$\times$ 1.4 N.A. objective and recorded using a monochrome CMOS camera at 500 fps. In this case, the Janus particle spontaneously crossed the center of the optical trap (marked by a red cross) and remained bound near the central region, exploring a three-dimensional volume. A randomly changing $z$-position of the particle was evident from the spontaneous change in its appearance as it moved above and below the imaging plane, which was set to coincide with the trapping plane. (playback speed: 1$\times$, duration: \SI{50}{\s})\\

\noindent
\href{https://drive.google.com/file/d/1FVDmsAIHrfFN67ccTtwQ7vAp020AFUFm/view?usp=drive_link}{\textbf{Supp. Video 2}}

\vspace*{0.1cm}
\noindent
Active dynamics of a Pt-silica Janus colloid (diameter $2a$ = \SI {1.76}{\micro \meter}) in a dilute $\mathrm{H}_{2} \mathrm{O}_{2}$ solution with concentration $c$ = 2.5\% (vol/vol) in an optical trap at laser power $P$ = \SI{7.62}{\mW} (Fig. 2(d)). Both diffusiophoresis and thermophoresis contributed to the self-propulsion of the Janus particle in this case. In this setting, the propulsion speed is $V$ = \SI {2.69}{\micro \meter /s}, and the ratio of the persistence time ($\tau_{\mathrm{R}}$) and equilibration time ($\tau_k$), $\tau_{\mathrm{R}}/\tau_k$ = 5.14. Here, the Janus particle moved freely along a circular path, maintaining a uniform distance from the center of the optical trap, which is indicated by a red cross. The particle remained below the focal point of the trap, which coincided with the imaging plane; hence, it appeared to be out-of-focus to some extent. The dynamics was observed under a microscope with a 60$\times$ 1.4 N.A. objective and recorded using a monochrome CMOS camera at 100 fps. (playback speed: 1$\times$, duration: \SI{50}{\s})\\

\noindent
\href{https://drive.google.com/file/d/1RNH7BM5JcLSk9Q3rXtX4Jz0FPxZIqfZb/view?usp=drive_link}{\textbf{Supp. Video 3}}

\vspace*{0.1cm}
\noindent
Numerically simulated dynamics of a harmonically bound active Brownian particle in two dimensions with propulsion speed $V$ = \SI {10}{\micro \meter /s}, persistence time $\tau_{\mathrm{R}}$ = \SI{1}{\s}, equilibration time $\tau_k$ = \SI{100}{\s}, and ratio of characteristic timescales $\tau_{\mathrm{R}}/\tau_k$ = 0.01 (Fig. 3(a)). A magnified view of the active Brownian particle is presented as a silica Janus colloid (green-grey) with Pt-coating (grey) on one hemisphere. The propulsion direction, from the coated (grey) to the uncoated (green) side, indicated by a black arrow, evolves with the orientational diffusion of the Janus particle. A radially symmetric harmonic confining potential centered at (0, 0) is represented by an orange color gradient. Starting from the center, the particle exhibits bound dynamics, where the instantaneous direction of the resultant or residual velocity $\boldsymbol{v_{\mathrm{res}}}$ (Eq. 1) is shown by a red arrow. Simulation time-step = \SI{1}{\ms}. (playback speed: 1$\times$)\\

\noindent
\href{https://drive.google.com/file/d/16bpfX74Cn7hS5IDzzbZ9Xkn4gPCN1FPX/view?usp=drive_link}{\textbf{Supp. Video 4}}

\vspace*{0.1cm}
\noindent
Numerically simulated dynamics of a harmonically bound active Brownian particle in two dimensions with propulsion speed $V$ = \SI {10}{\micro \meter /s}, persistence time $\tau_{\mathrm{R}}$ = \SI{100}{\s}, equilibration time $\tau_k$ = \SI{1}{\s}, and ratio of the characteristic timescales $\tau_{\mathrm{R}}/\tau_k$ = 100 (Fig. 3(d)). The active Brownian particle is shown as a silica Janus microsphere (green-grey) with a Pt-coating (grey) on one hemisphere. While the propulsion speed ($V$) remains constant, its direction, from the coated (grey) to the uncoated (green) side, indicated by a black arrow, evolves with the orientational diffusion of the Janus particle. The orange gradient represents the radially symmetric harmonic confining potential centered at (0, 0). Starting from the center, the particle mostly follows the initial propulsion direction, which is along the $x$-axis, until the propulsion changes its direction with slow orientational diffusion of the particle. The instantaneous direction of the resultant or residual velocity $\boldsymbol{v_{\mathrm{res}}}$ (Eq. 1) is indicated by a red arrow. The propulsion is counteracted by the restoring force (${-} k\boldsymbol{r}$) and eventually balanced at a radial distance $r = V \tau_k$ (marked by a dashed circle). Thus, the particle remains confined along the radial direction at $r = V \tau_k $, while performing a free and slow dynamics along the azimuthal direction, as the propulsion mostly points radially outwards. Simulation time-step = \SI{1}{\ms}. (playback speed: 1$\times$)\\